\documentclass[prd,nofootinbib]{revtex4}

\usepackage{latexsym}
\usepackage{epsfig}
\usepackage{amssymb}

\newcommand{\bea}{\begin{eqnarray}}
\newcommand{\eea}{\end{eqnarray}}
\newcommand{\be}{\begin{equation}}
\newcommand{\ee}{\end{equation}}

\begin{document}

\title{Cosmological perturbations in the Palatini formulation of
modified gravity}
\date{\today}

\author{Tomi Koivisto}
\email{tomikoiv@pcu.helsinki.fi}
\affiliation{Helsinki Institute of Physics,FIN-00014 Helsinki, Finland}
\author{Hannu Kurki-Suonio}
\email{hkurkisu@pcu.helsinki.fi}
\affiliation{Department of Physical Sciences, University of Helsinki, FIN-00014 Helsinki, Finland}

\begin{abstract}

Cosmology in extended theories of gravity is considered assuming the
Palatini variational principle, for which the metric and connection
are independent variables. The field equations are derived to linear
order in perturbations about the homogeneous and isotropic but
possibly spatially curved background. The results are presented in a
unified form applicable to a broad class of gravity theories allowing
arbitrary scalar-tensor couplings and nonlinear dependence on the
Ricci scalar in the gravitational action. The gauge-ready formalism
exploited here makes it possible to obtain the equations immediately
in any of the commonly used gauges. Of the three type of
perturbations, the main attention is on the scalar modes responsible
for the cosmic large-scale structure. Evolution equations are derived
for perturbations in a late universe filled with cold dark matter and
accelerated by curvature corrections. Such corrections are found to
induce effective pressure gradients which are problematical in the
formation of large-scale structure. This is demonstrated by analytic
solutions in a particular case. A physical equivalence between
scalar-tensor theories in metric and in Palatini formalisms is
pointed out.

\end{abstract}

\maketitle

\section{Introduction}

The observed acceleration of the
universe\cite{Knop:2003iy,Riess:2004nr,
Nesseris:2004wj,Spergel:2003cb} might indicate that our understanding
of gravity breaks down at cosmological distances. In the context of
general relativity, what is needed is some sort of fluid with large
negative pressure to accelerate the expansion. In its simplest form
this fluid is the cosmological constant\cite{Carroll:2000fy},
equivalent to a time-independent vacuum energy density. But there is
a problem with the cosmological constant, namely that according to
present knowledge of quantum field theory, one would expect this
vacuum energy density to be larger than observed by an enormous
factor.  One approach to this problem is to assume that due to a
symmetry the cosmological constant is zero, and in its stead a
dynamical fluid, generically called dark energy\cite{Peebles:2002gy},
provides the negative pressure. A plethora of different dark energy
models have been proposed, but none without some problems of its
own.

As an alternative to energetics of unknown fluids, it has been
proposed that the cosmic speed-up could stem from modifications to
general relativity\cite{Dvali:2000hr,Arkani-Hamed:2002fu,Sahni:2002dx,
Dvali:2003rk,Cline:2002mw}. Such modifications can come into
play when the gravitational action contains other curvature
invariants apart from the standard Einstein-Hilbert term. Then
general relativity can be considered as a limit of a hypothetical
more general theory. In fact suggestions for such a theory can be
found from fundamental physics. Quantization on curved spacetimes has
been found long ago to require extension of the Einstein-Hilbert
scheme by addition of higher-order curvature
terms\cite{Birrell:1982ix}, and it has been also shown that the
dominance of these terms at high energies could be the cause of the
early inflationary period in the
universe\cite{Starobinsky:1980te,Mukhanov:1990me,Nojiri:2003ft}. 
Similarly,
corrective terms that become important at small curvature can lead to
late-time effects seen in the cosmic expansion as an effective dark
energy\cite{Capozziello:2003tk,Carroll:2003wy,Nojiri:2003ni,
Nojiri:2004fw,Nojiri:2005jg,Brevik:2005ue,Multamaki:2005zs}. It 
has been shown that such curvature terms can
arise from compactification of time-dependent
extra dimensions in string/M-theory\cite{Nojiri:2003rz}.

Once the gravitational action is nonlinear in $R$, the question which
variational principle to apply becomes relevant. The Palatini variation
of a nonlinear gravity action leads to a different theory than the metric 
variation. The metric variation of extended gravity theories result in 
fourth order differential
equations which are difficult to analyze in practice. The Palatini
formulation, in which the connection is treated as an independent
variable\cite{Vollick:2003aw} is more tractable than the metric one,
and it can also in general exhibit better stability
properties\cite{Meng:2003ry,Meng:2003uv}, since it yields the
modified field equations as a second-order differential system.
Mathematical convenience does not of course prove that the Palatini
variation would be the fundamentally correct procedure. However, this
possibility might be interesting also according to some theoretical
prejudices. The second-order nature of the Palatini formulation is
conceptually more reconcilable with better-known physics than the
metric alternative, where the action in the beginning contains second
derivatives of the metric, and in the end one has to specify initial
values up to third derivatives to predict the evolution of the
system. The doubling of the variational degrees of freedom in the
Palatini formulation has an analogy with the Hamiltonian mechanics,
where the coordinates and momenta of particles are treated as
independent variables\cite{Misner}. On the more speculative side, it
is also interesting that the Palatini scheme of gravity can be
recovered in unification of general relativity with topological
quantum field theory\cite{Smolin:2003qu}.

Nonlinear Palatini gravity may also pass the Solar system
tests\cite{Meng:2003sx}. While this result has been
questioned\cite{Dominguez:2004ds,Olmo:2005hd},
the most recent discussions support
it\cite{Allemandi:2005tg,Sotiriou:2005xe,Sotiriou:2005hu}.
Cosmological background
solutions have been given for several choices of the gravitational
Lagrangian in the Palatini formalism, including (with some constants 
$\alpha$ and $\beta$) $f(R)=R^\alpha$\cite{Allemandi:2004ca}, $f(R)=\alpha \log(\beta
R)$\cite{Meng:2003en}, $f(R)=R+\alpha R^{-\beta}$\cite{Vollick:2003ic,
Meng:2003uv},
$f(R)=R + \alpha R^2$\cite{Meng:2003bk}, $f(R) = R + \alpha/R + \beta 
R^2$\cite{Meng:2004wg} and $f(R)=R+\alpha/R + \beta 
R^3$\cite{Sotiriou:2005hu}. Such solutions
have been tested against observational data, confirming that the late
accelerating expansion history can be produced viably in these
models\cite{Meng:2003uv,Kremer:2004bf,Capozziello:2004vh}. This
suggests that one should investigate the cosmological implications of
these theories beyond the background order. While a wide variety of
models is able to produce a scale factor evolution consistent with
observations, considerations of the cosmic microwave background
anisotropies and large scale inhomogeneities in cosmic structure may
distinguish between different physical assumptions underlying similar
or even identical expansion
histories\cite{Lue:2003ky,Koivisto:2004ne,
Koivisto:2005nr,Linder:2005in}. Recently it has been shown that early
inflation might also be modelled by nonlinear gravity in the Palatini
approach\cite{Sotiriou:2005hu}, and it will be interesting
to see the primordial perturbation spectra such models predict.

The aim of this paper is to set up perturbation equations for
generalized gravity in the Palatini approach. Cosmological
perturbation theory\cite{Bardeen:1980kt,Ellis:1989jt,Ellis:1989ju} of
modified gravities in the metric approach has been rigorously
formulated in the gauge-ready form by
Hwang and Noh\cite{Hwang:1990re,Hwang:1991aj,Hwang:1996xh}. Recent 
extensions
of these formulations encompass also kinetic
theory\cite{Hwang:2001qk}, tachyon condensation\cite{Hwang:2002fp}
and string corrections\cite{Hwang:2005hb}. The pioneering works found
in these references provide an arsenal of efficient methods on which
we draw to derive and analyze the Palatini versions of the perturbed
field equations. In particular, we will exploit the gauge-ready
formalism where the equations are instantaneously obtainable in
any of the convenient gauges employed for cosmological perturbations
in the literature. Since in the general case it is not clear
{\it a priori} which gauge choice is the most suitable to simplify
the analysis, leaving the temporal gauge unfixed in the
equations gives them very advantageous adaptability to different
situations.

In Section II we will review briefly the Palatini variational
principle. We also point out an equivalence between metric and
Palatini formulations of scalar-tensor theories, which is verified in
detail in Appendix A. In Section III we identify the degrees of
freedom in the cosmological metric and the energy-momentum tensor,
and derive the corresponding field equations for the three types of
perturbations, scalar, vector, and tensor. In Section IV we discuss
structure formation in $f(R)$ cosmologies, in particular curvature
corrections that become important in the late matter-dominated
universe. Section V contains discussion of our results. 

\section{Palatini approach to generalized gravity}

We consider gravity theories represented by the action
 \be \label{action_p}
 S = \int d^nx \sqrt{-g}
    \left[\frac{1}{2}f(R(g_{\mu\nu},\hat{\Gamma}^\alpha_{\beta\gamma}),\phi)
    - \frac{1}{2}\omega(\phi)(\partial\phi)^2 - V(\phi)
    + \mathcal{L}_m(g_{\mu\nu},\Phi,...)\right].
 \ee
Here $\Phi,...$ are some matter fields. In the Palatini approach one
lets the torsionless connection $\hat{\Gamma}^\alpha_{\beta\gamma}$
vary independently of the metric. The Ricci tensor is then defined
solely by this connection,
 \be \label{ricci}
 R_{\mu\nu} \equiv \hat{\Gamma}^\alpha_{\mu\nu , \alpha}
       - \hat{\Gamma}^\alpha_{\mu\alpha , \nu}
       + \hat{\Gamma}^\alpha_{\alpha\lambda}\hat{\Gamma}^\lambda_{\mu\nu}
        - \hat{\Gamma}^\alpha_{\mu\lambda}\hat{\Gamma}^\lambda_{\alpha\nu},
 \ee
whereas the scalar curvature is given by
 \be \label{ricci_s}
    R \equiv g^{\mu\nu}R_{\mu\nu} \,.
 \ee
The matter energy-momentum tensor is defined as
 \be \label{memt}
 T_{\mu\nu}^{(m)} \equiv -\frac{2}{\sqrt{-g}} \frac{\delta
 (\sqrt{-g}\mathcal{L}_m)}{\delta(g^{\mu\nu})}.
 \ee
Hereafter we drop the sub- and superscripts $m$, which stand
collectively for all matter except the scalar field $\phi$. We
consider the second and the third term in the action (\ref{action_p})
as the Lagrangian for $\phi$. Then the scalar field energy-momentum
tensor
 \be \label{scalaremt}
    T^{\mu (\phi)}_\nu = \omega(\nabla^
   \mu\phi)(\nabla_\nu\phi) - \delta^\mu_\nu\left[\frac{1}{2}\omega
   (\partial\phi)^2 + V\right].
 \ee
is found by a similar variation as Eq.(\ref{memt}). The field
equations which follow from extremization of the action,
Eq.(\ref{action_p}), with respect to metric variations, can be
written as
 \be \label{fields2} F R^\mu_\nu -\frac{1}{2}f\delta^\mu_\nu =
 T^\mu_\nu + T^{\mu (\phi)}_\nu,
 \ee
where we have defined 
\be
F \equiv \partial f/\partial R.
\ee
In general relativity, $f = R/8\pi G$ or $(R-2\Lambda)/8\pi G$, and $F = 
1/8\pi G$.

Note that the covariant derivative $\nabla$ in Eq.~(\ref{scalaremt})
is taken using the Levi-Civita connection (i.e., the Christoffel
symbol) of $g_{\mu\nu}$,
 \be \label{gcon}
    \Gamma^\alpha_{\beta\gamma} \equiv \frac12\,g^{\alpha\lambda}
    \left(g_{\lambda\beta,\gamma} + g_{\lambda\gamma,\beta}-
    g_{\beta\gamma,\lambda}\right).
 \ee
Thus we have two different connections, the ``Palatini'' connection
$\hat{\Gamma}^\alpha_{\beta\gamma}$, appearing as an independent
variable in the action (\ref{action_p}), and the Levi-Civita
connection $\Gamma^\alpha_{\beta\gamma}$ derived from the metric
$g_{\mu\nu}$, and two different covariant derivatives, $\hat{\nabla}$
and $\nabla$, corresponding to these two connections. We denote by
$R_{\mu\nu}(g) \equiv \Gamma^\alpha_{\mu\nu , \alpha}
       - \Gamma^\alpha_{\mu\alpha , \nu}
       + \Gamma^\alpha_{\alpha\lambda}\Gamma^\lambda_{\mu\nu}
        - \Gamma^\alpha_{\mu\lambda}\Gamma^\lambda_{\alpha\nu}$
the Ricci tensor constructed from the Levi-Civita connection to
distinguish it from the Ricci tensor $R_{\mu\nu}$ of
Eq.~(\ref{ricci}). Likewise, we write $R(g) \equiv
g^{\mu\nu}R_{\mu\nu}(g)$ to distinguish this scalar curvature,
derived solely from the metric, from the scalar curvature of
Eq.(\ref{ricci_s}) appearing in the action (\ref{action_p}).

By varying the action with respect to
$\hat{\Gamma}^\alpha_{\beta\gamma}$, one gets the condition
 \be
 \hat{\nabla}_\alpha\left[\sqrt{-g}g^{\beta\gamma}F\right]=0,
 \ee
implying that this connection is compatible with the conformal metric
 \be
 \hat{g}_{\mu\nu} \equiv F^{2/(n-2)}g_{\mu\nu}.
 \ee
This connection governs how the tensor $R_{\mu\nu}$ appearing in the
action settles itself in order to minimize the action, but it turns
out that the connection of Eq.(\ref{gcon}) determines the geodesics
that freely falling particles follow, since energy momentum is
conserved according to this connection\cite{Koivisto:2005yk},
 \be \label{em}
 \nabla_\mu T^\mu_\nu = 0,
 \ee
whereas in general $\hat{\nabla}_\mu T^\mu_\nu \neq 0$. Therefore one
may interpret the Levi-Civita connection as the gravitational field
as in Ref.\cite{Magnano:1995pv}. In any case it is convenient to
write the modified field equations in terms of the connection
$\Gamma$. One finds that the Ricci tensor is
 \be \label{riccit}
 R_{\mu\nu} = R_{\mu\nu}(g) +\frac{(n-1)}{(n-2)}
             \frac{1}{F^2}(\nabla_\mu F)(\nabla_\nu F) -
             \frac{1}{F}(\nabla_\mu \nabla_\nu F)
            -\frac{1}{(n-2)}\frac{1}{F}g_{\mu\nu}\Box F.
 \ee
The curvature scalar and Einstein tensor follow straightforwardly,
 \be \label{riccis}
 R = R(g) + \frac{n-1}{(n-2)F}\left[-2 \Box F
    + \frac{1}{F}(\partial F)^2\right],
 \ee
 \be
 G_{\mu\nu} =  G_{\mu\nu}(g) +
             \frac{(n-1)}{(n-2)}\frac{1}{F^2}(\nabla_\mu
               F)(\nabla_\nu F)
             - \frac{1}{F}\left(\nabla_\mu \nabla_\nu  -
               g_{\mu\nu}\Box\right)F
             -\frac{(n-1)}{2(n-2)}\frac{1}{F^2}g_{\mu\nu} (\partial
              F) ^2.
 \ee
From now on we set the spacetime dimension to $n=4$ and use units
$8 \pi G \equiv c \equiv 1$. One useful way to write the field
equations is in the form of Einstein gravity plus corrections:
 \bea \label{eg} G^\mu_\nu(g) & = & {T^\mu_\nu} +
 {T^\mu_\nu}^{(\phi)}
               + (1-F)R^\mu_\nu(g) \nonumber \\
             & - & \frac{3}{2F}(\nabla^\mu F)(\nabla_\nu F)
               + (\nabla^\mu \nabla_\nu F)
               + \frac{1}{2}\left[(f-R) + (1-\frac{3}{F})\Box F +
     \frac{3}{2F}(\partial F)^2\right]\delta^\mu_\nu.
 \eea
Due to Eq.(\ref{em}) and the Bianchi identity, the correction terms
are covariantly conserved according to $\Gamma$. In the metric
formalism, analogous corrections appear even in vacuum, and it is
natural to treat them as an effective additional
fluid\cite{Hwang:1990re}. However, here these corrections can be
expressed as functions of the trace of the matter energy-momentum
tensor $T \equiv g^{\mu\nu}T_{\mu\nu}$ and the scalar field $\phi$
and its derivatives, and one can view Eq.(\ref{eg}) as general
relativity with nonstandard matter couplings. Thus the whole
right-hand side may be regarded as an effective matter
energy-momentum tensor. In vacuum it reduces to a cosmological
constant\cite{Ferraris:1992dx}, and in the case of conformal matter
(i.e. $T=0$ \cite{Alnes:2005ed}) to the usual $T_{\mu\nu}$. One finds
$R$ in terms of matter from the trace of the field equation,
Eq.(\ref{fields2}),
 \be \label{trace} FR-2f = \omega (\partial
 \phi)^2 - 4 V + T.
 \ee
We will refer to this central relation as the structural
equation\cite{Allemandi:2004yx,Allemandi:2005qs}.

Now consider some specific cases. Perhaps the simplest nontrivial example
is a pure scalar-tensor theory, for which $f=F(\phi)R$. Then the
structural equation becomes
 \be \label{st_trace}
 R = -\frac{\omega \dot{\phi}^2 - 4V + T}{F}.
 \ee
We remark here that this theory is physically equivalent to a similar
one in the metric formulation with a modified kinetic term of the
scalar field. Namely, with arbitrary $F(\phi)$ and $\omega(\phi)$,
the kinetic coefficient $\omega_\mathrm{metric}$ for the scalar field
in the corresponding metric formulation of scalar-tensor theory is
 \be \label{correspondence}
 \omega_\mathrm{metric}(\phi) = \omega(\phi) - \frac{3 F'^2(\phi)}{2F(\phi)}.
 \ee
We show in Appendix A that the metric and Palatini
approaches yield exactly the same field equations and equations of
motion when this rescaling of $\omega$ is taken into account.
Therefore no qualitatively new features, regardless of the scalar
field couplings, are introduced in the Palatini approach whenever the
gravitational action is linear in $R$.
This correspondence, Eq.(\ref{correspondence}), enables one to extend
the results of studies of scalar-tensor gravity\cite{Fujii:2003pa}
within the metric formalism to the Palatini framework, and may
motivate some previously less-investigated forms of these theories.

Consider the case without a non-minimally coupled scalar field, $f =
f(R)$. We assume that there exists a solution $R=R(T)$ to the
structural equation $FR-2f=T$. For example, if $f = R-\alpha_0/(3R)$,
then \be \label{oneperr} R(T) = -\frac{1}{2}\left(T + \sqrt{T^2 +
4\alpha_0}\right). \ee If $f(R) \sim  R + \alpha_0 R^2$, we have
$R(T)=-T$ as in general relativity. If $f=R^2$, the theory admits
only vacuum or conformal matter.

\section{Cosmological equations}

\subsection{Definitions}

The line-element in the perturbed Friedmann-Robertson-Walker (FRW)
spacetime can be written as (see e.g. Ref.\cite{Hwang:2001qk} for
an only slightly different notation)
 \be \label{metric}
  ds^2 = a^2(\eta)\left\{-\left(1+2\alpha\right)d\eta^2 - 2\left(\beta_{,i}
       + b_i\right)d\eta dx^i +
       \left[ g^{(3)}_{ij}+2\left(g^{(3)}_{ij}\varphi + \gamma_{|ij} +
              c_{(i|j)} + h_{ij}\right) \right]dx^idx^j\right\}.
 \ee
We characterize the scalar perturbations in a general gauge by the
four variables $\alpha,\beta,\varphi,\gamma$. Vector perturbations
introduce four more degrees of freedom, the divergenceless 3-vector
fields $b_i$ and $c_i$. Gravitational waves are described by the two
free components of the symmetric, transverse and traceless 3-tensor
$h_{ij}$. We have thus decomposed the ten independent components of
the symmetric tensor $\delta g_{\mu\nu}$ into three types of
perturbations according to their transformation properties under
spatial rotations. The comoving spatial background metric
$g^{(3)}_{ij}$ reduces to $\delta_{ij}$ in a flat universe. The
vertical bar indicates a covariant derivative based on the
Levi-Civita connection of $g^{(3)}_{ij}$. This metric is used to
lower and raise spatial indices $i,j,k \dots$ of the perturbation
variables.

The components of the energy-momentum tensor for an imperfect fluid
are
 \be \label{fluid}
 T^0_0 = -(\bar{\rho}+\delta\rho), \quad
 T^0_i=-\left(\bar{\rho}+\bar{p}\right)\left(v{,_i} +
 v^{(v)}_i\right), \quad T^i_j = (\bar{p}+\delta p)\delta^i_j +
 \Pi^i_j.
 \ee
Here $\rho$ and $p$ are energy density and pressure, and $v$,
$v^{(v)}$ are the scalar and vector velocity perturbations\footnote{Note 
that we use the covariant velocity perturbations\cite{Malik:2004tf}, 
sometimes denoted as $v \equiv a(V-\beta)$, and $v^{(v)}_i 
\equiv a(V^{(v)}_i-b_i)$.}, respectively. Background quantities are denoted 
with an overbar, which we will usually omit when unnecessary. The isotropy 
of background does not allow anisotropic stress except as a
perturbation. This we decompose into the scalar, vector and tensor
contributions as
 \be \Pi_{ij} \equiv \left(\Pi^{(s)}_{|ij}
         + \frac{1}{3}\triangle\Pi^{(s)} \right) +
         \Pi^{(v)}_{(i|j)} +
         \Pi^{(t)}_{ij},
 \ee
where $\triangle$ stands for the three-space Laplacian based on the
Levi-Civita connection of $g^{(3)}_{ij}$. The vector $\Pi^{(v)}_i$ is
divergenceless and the tensor $\Pi^{(t)}_{ij}$ is symmetric,
transverse, and traceless. The four scalar degrees of freedom for the
fluid perturbation are therefore $\delta$, $\delta p$, $v$, and
$\Pi^{(s)}$, independent components of the divergenceless vectors
$v^{(v)}_i$ and $\Pi^{(v)}_i$ sum up to four and the tensor
describing gravitational waves $\Pi^{(t)}_{ij}$ has two independent
components.

Some of these degrees of freedom are due to arbitrariness in
separating the background from the perturbations. In the gauge-ready
formalism\cite{Hwang:1991aj} one deals with these gauge degrees of
freedom by noting that the homogeneity and isotropy of the background
space implies invariance of all physical quantities under purely
spatial gauge transformations. Therefore one can trade $\beta$ and
$\gamma$ to the shear perturbation
 \be \label{gready}
 \chi \equiv a(\beta+\dot{\gamma}) \,.
 \ee
where an overdot means derivative with respect to the conformal time
$\eta$. Since both $\beta$ and $\gamma$ vary under spatial gauge
transformation, they appear only through the spatially invariant
linear combination $\chi$ in all relevant equations. In addition, one
can define the perturbed expansion scalar
 \be
 \kappa \equiv \frac{3}{a}(H\alpha -
 \dot{\varphi})-\frac{\triangle}{a^2}\chi\,,
 \ee
Here $H$ is the Hubble parameter defined with respect to conformal
time, i.e, the usual Hubble parameter multiplied by the scale factor
$a$. The variable $\kappa$ is a convenient linear combination, the
use of which simplifies some equations, but it is not linearly
independent of other perturbations. Only three of the variables
 $\alpha$, $\varphi$, $\chi$ and $\kappa$ are independent.
All of them can be physically interpreted as perturbations of the
normal-frame vector\cite{Hwang:1991aj}. The advantage of using this
set of variables is based on the fact that they are spatially
gauge-invariant. Thus, writing equations in terms of them, one can
conveniently fix the temporal gauge by just setting one of these
metric perturbations to zero. The synchronous gauge, corresponding to
$\alpha = 0$, is an exception where the gauge mode is removed only up
to a constant. For example, the longitudinal (also called the
Newtonian or the zero-shear) gauge is the one where $\chi=0$. One
popular gauge is the comoving one, where, instead of setting any of
the metric perturbations to zero, one sets the fluid velocity
perturbation $v$ in Eq.(\ref{fluid}) to $v=0$ . Suitable linear
combinations of the above gauge conditions can be considered also.

Similarly, we will exploit the spatially gauge-invariant variable
 \be
    \Psi_i \equiv b_i + \dot{c}_i
 \ee
to characterize vector perturbations of the metric. The tensor perturbation
$h_{ij}$ is gauge-invariant by construction.

\subsection{Background}

The modified Friedmann equation corresponding to Eq.(\ref{fields2}) is
 \be \label{friedmann}
 3H^2 = \frac{1}{F}\left[a^2\rho +\frac{\omega}{2}\dot{\phi}^2 + a^2V -
       3H\dot{F} -
       \frac{3}{4F}\dot{F}^2 -
       \frac{a^2}{2}\left(f-FR\right)\right] - 3K,
 \ee
where $K$ is the curvature of the background space.
One should be equipped with a solution to the structural
equation Eq.(\ref{trace}), with which to replace the scalar curvature
$R$ appearing both explicitly and implicitly (through $f$, $F$ and
their time derivatives) in Eq.(\ref{friedmann}). For a review of
background solutions in several $f(R)$ cases, see\cite{Wang:2004vs}.
An implicit expression for $R$ follows from Eq.(\ref{riccis}):
 \be
 a^2 R = 6\left(\dot{H}+H^2+K\right) + \frac{3}{F}\left(\ddot{F}
 + 2H\dot{F} - \frac{1}{2F}\dot{F}^2\right).
 \ee

The matter continuity equation is, as usual,
 \be
 \dot{\rho} + 3H(\rho + p)=0.
 \label{cont1}
 \ee
Individual matter species, if minimally
coupled to gravity and without interactions with other matter species, also
satisfy this equation. The Klein-Gordon equation for the scalar field is
 \be
    \ddot{\phi}+2H\dot{\phi} +
    \frac{\omega'}{2\omega}\dot{\phi}^2 + \frac{1}{2\omega}\left(2V - f
    \right)_{,\phi} = 0.
 \label{cont2}
 \ee

\subsection{Field equations for scalar perturbations}

First we present the equations that govern the evolution of
scalar perturbations defined in the subsection IIIA. 
From now on we consider variables in the Fourier
space. The transformation is simple since at linear order each
$k$-mode evolves independently.

The energy constraint ($G^0_0$ component of the field equation) in
generalized gravity is
 \bea
 \left(2H+\frac{\dot{F}}{F}\right)a\kappa & + &
 \left(6K-2k^2\right)\varphi + \frac{1}{F}\left(-\omega \dot{\phi}^2
 + \frac{3\dot{F}^2}{2F} + 3H\dot{F}\right)\alpha
 \nonumber \\
 & = & \frac{1}{F}\Bigg\{-a^2\delta\rho - \omega\dot{\phi}\delta\dot{\phi}
 - \frac{1}{2}\left[\omega'\dot{\phi}^2 + a^2\left(2V - f \right)_{,\phi}\right]\delta \phi
 \nonumber \\
 &  & +
 \left[
 3\left(H^2+K^2\right)-\frac{3\dot{F}^2}{4F^2}-\frac{a^2}{2}R +
 k^2\right]\delta F +
 \left(\frac{3}{2}\frac{\dot{F}}{F}+3H\right)\dot{\delta F} \Bigg\},
 \label{admenergy}
 \eea
and the momentum constraint ($G^0_i$ component) is
 \be
  a\kappa - \left(k^2-3K\right)\frac{1}{a}\chi +
 \frac{3\dot{F}}{2F}\alpha = \frac{3}{2F}\left\{
 a^2\left(\rho+p\right)\frac{v}{k} + \omega\dot{\phi}\delta\phi
 -\left(H+\frac{3\dot{F}}{2F}\right)\delta F  + \dot{\delta
 F}\right\}.
 \label{momentum}
 \ee
The shear propagation equation ($G^i_j-\frac{1}{3}\delta^i_jG^k_k$
component) reads
 \be
 \frac{1}{a}\dot{\chi}+\left(H+\frac{\dot{F}}{F}\right)\frac{1}{a}\chi
 -\alpha - \varphi = \frac{1}{F}\left(a^2\Pi^{(s)} + \delta F\right).
 \label{propagation}
 \ee
The Raychaudhuri equation ($G^k_k-G^0_0$ component) is now given by
 \bea
 2a\dot{\kappa} & + &
 \left(4H+\frac{\dot{F}}{F}\right)a\kappa +
 \left[6\left(\dot{H}-H^2\right) + 6\left(\frac{\ddot{F}}{F} -
 \frac{\dot{F}^2}{F^2}\right) -3H\frac{\dot{F}}{F} + \frac{4}{F}\omega
 \dot{\phi}^2 - 2k^2\right]\alpha + 3\frac{\dot{F}}{F}\dot{\alpha}
 \nonumber \\
 & = & \frac{1}{F}\Bigg\{ a^2\left(\delta \rho + 3\delta p\right) +
 4\omega\dot{\phi}\delta\dot{\phi} +
 \left[2\omega' \dot{\phi}^2 + a^2\left(f-2V\right)_{,\phi}\right]\delta\phi
 \nonumber \\
 &  & + \left(
  3\frac{\dot{F}^2}{F^2} - a^2R + 6\dot{H} + k^2\right)\delta F
 -6\frac{\dot{F}}{F}\dot{\delta F} + 3\ddot{\delta F}\Bigg\}.
 \label{ray}
 \eea
The scalar curvature perturbation can be found by combining previous
equations or directly by using Eq.(\ref{riccis}),
 \bea
 a^2\delta R & = & -2a\dot{\kappa} - \left(8H+3\frac{\dot{F}}{F}\right)a\kappa
 +3\left[2\left(H^2-\dot{H}-\frac{\ddot{F}}{F}\right) + \frac{\dot{F}^2}{F^2}
 -\frac{\dot{F}}{F}H + \frac{2}{3}k^2\right]\alpha +
 2\left(2k^2-6K\right)\varphi - 3\frac{\dot{F}}{F}\dot{\alpha}
 \nonumber \\
 & + & \frac{3}{F}\left[\left(
  \frac{\dot{F}^2}{F^2} -2H\frac{\dot{F}}{F} -\frac{\ddot{F}}{F}
  + k^2\right)\delta F
  +\left(2H-\frac{\dot{F}}{F}\right)\dot{\delta F} + \ddot{\delta F}\right].
 \eea
Also redundant with Eqs.(\ref{admenergy}--\ref{ray}) is the perturbed
version of the structural equation ($G^\alpha_\alpha$ component)
 \be
 R\delta F + F\delta R - 2\delta f =
 \frac{2}{a^2}\omega \dot{\phi}\left(\delta \dot{\phi} - \dot{\phi}\alpha\right)
 + 2(f-2V)_{,\phi} \delta\phi - \delta \rho + 3\delta p .
 \ee

We have $\delta f = F\delta R + f_{,\phi}\delta \phi$, and similarly
$\delta F = F_{,R}\delta R + F_{,\phi}\delta\phi$. However, these
functions can also be expressed in a form proportional to matter
perturbations using the structural equation Eq.(\ref{trace}).
Consider the earlier example $f = R-\alpha_0/(3R)$,
Eq.(\ref{oneperr}). There we have
 \be
 \delta f = \left(1+\frac{\alpha_0}{3R^2(T)}\right)\left(-\frac{1}{2}-
 \frac{T}{2\sqrt{T^2+4\alpha_0}}\right)\left(-\delta\rho+3\delta
 p\right),
 \ee
 \be \label{deltaf}
 \delta F =
 \frac{\alpha_0}{3R^3(T)}\left(1 +
 \frac{1}{\sqrt{T^2+4\alpha_0}}\right)\left(-\delta\rho+3\delta
 p\right).
 \ee
It is straightforward to obtain the derivatives $\dot{\delta F}$ and
$\ddot{\delta F}$ from Eq.(\ref{deltaf}). In this manner $\delta R$
can be related to $\delta T$, and one can consider the right-hand
sides of the perturbed field equations as matter sources in the
modified gravity. However, one possibly appealing gauge condition is
to set $\delta F = 0$. This eliminates the complicated effective
matter source terms in the right hand side, while leaving some
$f$-dependent terms to modify the evolution of the metric
perturbations. In the $f=f(R)$ case this is the gauge where $\delta T
=0$.

The equation of motion for the scalar field is the same as in the
metric formulation,
 \bea \label{klein}
 \ddot{\delta \phi} + \left( \frac{\omega '}{\omega}\dot{\phi} +
 2H\right)\dot{\delta \phi} & + & \left[k^2 +
 \frac{\omega ''}{2\omega}\dot{\phi}^2 + \frac{\omega '}{\omega}
 \left(\ddot{\phi}+2H\right) + \frac{1}{2\omega}\left(2V-f\right)_{,\phi\phi}
 \right]\delta\phi
 \nonumber \\
 & =  & \dot{\phi}\left( \dot{\alpha} + a\kappa\right)
 + \left(2\ddot{\phi}+H\dot{\phi} + 2\frac{\omega '}{\omega}\dot{\phi}^2\right)\alpha
 +\frac{1}{2\omega}F_{,\phi}\delta R.
 \eea
However, now the interpretation of $R$ appearing on both sides of
Eq.(\ref{klein}) is different. The minimally coupled individual matter
species obey the standard continuity and Euler equations,
 \be \label{pcont1}
 \delta \dot{\rho} + 3H\left(\delta\rho+\delta p\right) =
 (\rho+p)\left(-kv+a\kappa -3H\alpha\right),
 \ee
 \be \label{pcont2}
 \frac{1}{a^4}\left[a^4(\rho+p)v\right]^{.}=k\left(\rho+p\right)\alpha +
 k\delta p
 - \frac{2}{3}\frac{k^2-3K}{k}\Pi^{(s)},
 \ee
since matter fields in the action, Eq.(\ref{action_p}), decouple from
the connection.

\subsection{Vector and tensor perturbations}

Vector and tensor modes are simpler in an FRW spacetime than scalar
perturbations. This is true especially in $f(R,\phi)$ gravity, since
neither $R$ nor $\phi$ has vector or tensor components, regardless of
whether we are in the Palatini or metric formulation. Therefore we
can employ the analogous results found in the metric case, see for
example\cite{Hwang:2001qk}. The matter quantities in this subsection
do not involve contributions from the scalar field, since it cannot
generate vector or tensor perturbations.

The equations governing the evolution of rotational perturbations are
 \be \label{rotfield}
    \frac{k^2 - 2K}{2a^2}\Psi_i =
     \frac{1}{F}(\rho+p)v^{(v)}_i,
 \ee
 \be \label{rotcons}
     \frac{1}{a^2}\left[a^4(\rho+p)
      v^{(v)}_i\right]^{.}= - \frac{k^2-2K}{2k^2}\Pi^{(v)}_i.
 \ee
The first is the field equation (the $G^0_i$ component), the second
one the conservation equation. The only difference to general
relativity here is the appearance of the prefactor $1/F$ modulating
the amplitude of $\Psi_i$. For example, one can see that the angular
momentum of a perfect fluid $\sim a^4(\rho+p)v^{(v)}_i$ is conserved
in an expanding universe.

The gravitational wave is governed by the equation
 \be
 \ddot{h}^i_j + \left(2H+\frac{\dot{F}}{F}\right)\dot{h}^i_j
 + \left(k^2+2K\right)h^i_j
 = \frac{a^2}{F}{\Pi^i_j}^{(t)}.
 \ee
Modifications to general relativity appear here as an additional
damping term and a similar modulating of the source term as for the
rotation. In the absence of anisotropic stresses, only the former has
an effect. For a flat universe, an integral solution exists for the
superhorizon scales,
 \be
   h^i_j(\eta,k) =  A^i_j - B^i_j \int^\eta\frac{1}{a^2F}d\eta,
 \ee
where $A^i_j$ and $B^i_j$ are constants for each $k$-mode. Thus only
the decaying solution is modified when $f \neq R$.

\section{Structure formation in $f(R)$ cosmologies}

\subsection{Inhomogeneities in late universe}

Now we are in a position to consider formation of structure in
cosmologies derived from the Palatini approach to generalized
gravity. As discussed in the introduction, it is of particular
interest to uncover whether curvature corrections driving the late
acceleration will predict testable features in the CMB or large scale
structure. Therefore we will consider $f(R)$ theories in a flat
universe dominated by pressureless cold dark matter. There is no dark 
energy, since we consider modified gravity as its alternative.  

The evolution of matter perturbations is governed by the equations
(\ref{pcont1}) and (\ref{pcont2}), which for pressureless
matter reduce to
 \be \label{cont3}
   \dot{\delta} = -kv+a\kappa-3H\alpha, \quad \dot{v} = -Hv + k\alpha.
 \ee
We will work first in the uniform density gauge (indicated by the
subscript ${}_\delta$), where $\delta_\delta=0$. Now also $\delta
f_\delta$ and $\delta F_\delta$ together with their derivatives
vanish, considerably simplifying the analysis. (This of course does not mean
that the matter perturbation disappears, it is just carried by other
perturbation quantities, as the physics is completely independent of gauge
choice.  We later transfer to the comoving gauge, where the interpretation
of the results is more straightforward.)

In the uniform density gauge, manipulation of the field equations 
(\ref{admenergy}--\ref{ray}) with the 
help of the conservation equations (\ref{cont3}) then yields an evolution
equation for the non-vanishing matter velocity perturbation
$v_\delta$,
 \be \label{vdelta}
   \ddot{v}_\delta =  \frac{1}{F(2FH+\dot{F})}\left\{
   \left[-2F^2(H^2+2\dot{H}) + 2\dot{F}^2-\dot{F}FH-2\ddot{F}F\right]\dot{v}_\delta -
   \left[6F^2\dot{H}H-2\dot{F}^2H+\dot{F}F(\dot{H}+\frac{k^2}{3})+
   2\ddot{F}FH\right]v_\delta\right\}.
 \ee
By solving this relatively simple (depending on the form of $f(R)$)
differential equation, one can easily find also the metric
perturbations, since they are related to $v_\delta$ and its
derivatives. These solutions can also be related to solutions in any
other gauge by using the gauge transformation properties of the
relevant variables. In fact, although the evolution equation
(\ref{vdelta}) is particularly tractable in the uniform-density
gauge, it is difficult to physically interpret the significance of
modifications to general relativity from this equation.

Therefore we transform this result into the more intuitive comoving
gauge (subscript ${}_v$), which is defined by $v_v=0$. We cannot of
course find a nontrivial equation for $v_v$, but will instead
consider the comoving gauge density perturbation $\delta_v$. By
looking at Eq.(\ref{cont3}) one sees that this gauge also coincides
with the synchronous one at the late matter-dominated stage of the
universe, regardless of the form of $f(R)$. By using their gauge
transformation properties, we find a convenient relation between the
perturbation variables in the different gauges,
 \be
 \delta_v = 3Hv_\delta/k.
 \ee
Using this we can recast Eq.(\ref{vdelta}) into the form
 \bea \label{dddelta}
 \ddot{\delta}_v & = & \frac{1}{3FH^2(2FH+\dot{F})}\Big\{-3H\left[
 2FH(FH^2+\ddot{F})  -
 2\dot{F}^2H+\dot{F}F(-2\dot{H}+H^2)\right]\dot{\delta}_v
 \nonumber \\ & + &
 \left[6F^2H^2(\ddot{H} - 2\dot{H}H) + 6\dot{F}^2H(H^2-\dot{H}) +
  \dot{F}F(3\ddot{H}H - 6\dot{H}^2 - H^2 k^2)
  + 6\ddot{F}FH(\dot{H} - H^2)\right]\delta_v \Big\}.
 \eea
We have checked that the form of this equation does not change when
the assumption $K = 0$ is relaxed. 

For the case of general relativity with a cosmological constant,
$f(R)=R - 2\Lambda$, Eq.(\ref{dddelta}) reduces to
 \be \label{delta_s0}
 \ddot{\delta}_v = -H\dot{\delta}_v + \left(\frac{\ddot{H}}{H}-2\dot{H}
 \right)\delta_v.
 \ee
If the cosmological constant and cold dark matter are thought of
as components of a two-component cosmic fluid, the equation of state
of this total fluid is $w_T = -(1 + a^{-3}\rho_0/\Lambda)^{-1}$. The
density contrast $\delta_v^T$ can then be written as
$\delta_{T} = (1+w_T)\delta_v$. Inserting these in Eq.(\ref{delta_s0})
gives the familiar
 \be \label{delta_s}
 \ddot{\delta}_{T} = (6w_T-1)H\dot{\delta}_{T} +
 \frac{3}{2}[1+8w_T-3w_T^2]H^2\delta_{T}.
 \ee
Notice that this equation is scale-independent, whereas
Eq.(\ref{dddelta}) involves a $k^2$ -term, which stays small only
at very large scales.

Consider first this large scale limit of Eq.(\ref{dddelta}). There
both the ``friction'' and the ``source'' term in Eq.(\ref{delta_s})
are non-trivially modified. Similar effects appear when the
cosmological constant is replaced by dynamical dark
energy\cite{Amarzguioui:2004kc,Koivisto:2005nr}. Here these modifications depend on the
specific form of $f(R)$, and one might recover an evolution of
$\delta_v$ leading to a large-scale matter power spectrum consistent
with observations, by choosing a suitable function of $R$.

However, there could still be difficulties with the
small-scale structure in these models. This is because the gradient term
in Eq.(\ref{dddelta}) acts as the
dominant source inside the horizon. 
The gradient appears since the curvature corrections 
induce effective pressure fluctuations in the inherently cold dark matter. 
This may have been anticipated from the 
field equation (\ref{eg}), where it is seen that the modified 
Palatini gravity is coupled to derivatives of matter energy momentum. 
Since the cosmological background is homogeneous, modifications to the 
Friedmann equation stem only from the time variation of the background 
density of matter. However, the evolution of inhomogeneities in the 
universe is inevitably affected by the response of the modified 
Palatini gravity to spatial variations in the distribution of matter. 
This is why we have found here an effective sound speed for cold dark 
matter.

It might be useful to compare the situation with dark energy models.
A cosmological constant does not fluctuate, therefore the absence 
of a gradient term in Eq.(\ref{delta_s}). On the other hand, quintessence 
fields fluctuate but their gradients are negligible, since those fields 
are very smooth inside horizon, although this may not be the case if 
the quintessence field is coupled with dark matter\cite{Koivisto:2005yk}. 
The models unifying dark energy and dark 
matter into a single component, are plagued by a nonzero sound speed at small 
scales\cite{Koivisto:2004ne,Amarzguioui:2004kc}. These unified models are 
somewhat analogous to the case studied here, since we assume that due to 
the properties of non-standard gravity, 
the accelerated expansion of the universe, usually explained by dark 
energy, is in fact driven only by dark matter. Stability of modified 
gravities in the metric approach \cite{Dolgov:2003px,Nojiri:2003wx} has 
been explored in de Sitter
space using covariant and gauge invariant analysis of cosmological
perturbations\cite{Faraoni:2004dn,Faraoni:2005ie}, but it remains to
be studied whether similar effects in the evolution of matter
inhomogeneities as seen here will appear in the metric approach.

Here, at sufficiently large $k$, the small scale evolution of
$\delta_v$ is determined by
 \be \label{gradient}
   \ddot{\delta}_v = -\frac{F'(T)\rho}{2F(T)+3F'(T)\rho}k^2\delta_v,
 \ee
where we have expressed the gradient in Eq.(\ref{dddelta}) in terms of the
background solution for $F$ one finds from the structural equation. It is 
interesting to note that the gradient (or the effective sound speed) does 
identically vanish only in the case of Einstein gravity with a 
cosmological constant. For example, when $f(R)=R-\alpha_0/(3R)$ one has
 \be \label{gradient2}
   \ddot{\delta}_v =
   -\left[\frac{2x}{3(1+6x)-(3+8x)\sqrt{1+4x}}\right]k^2\delta_v,
 \ee
where the square bracket term is of the order one when $x \equiv
\alpha_0\rho^{-2}$ is, which holds when the curvature and matter
energies are both relevant. Indeed the effective pressure fluctuation
seems to become troublesome inside the horizon.

\subsection{A particular example: $f(R) \sim R^n$}

We demonstrate the general considerations of the previous subsection
with a specific choice for the nonlinear Lagrangian, $f(R) \sim R^n$,
where $n \neq 0,2,3$\footnote{The limit $n=3/2$ appears
singular in our equations.
This is because at that limit the deceleration parameter vanishes and
the conformal Hubble parameter cannot be expressed in the form $H
\sim 1/\eta$. When $n<3/2$, $\eta$ is positive, and when $n>3/2$,
$\eta$ is growing from negative values towards zero. Our results are
still well defined at the limit $n=3/2$. On the other hand, $n=2$ is
a physically singular limit of the theory. This is not seen
explicitly in our equations, since factors like $1/(n-2)$ belong to
an otherwise irrelevant proportionality constant in the solution for
$f$ which we have rescaled away. We have also a singularity at n=3
\cite{Allemandi:2004ca}.}.
Such a case has been studied previously\cite{Allemandi:2004ca,
Allemandi:2004wn} and shown to predict a plausible background evolution 
at late times\cite{Capozziello:2004vh}. In fact the background is simply
described by a constant effective equation of state in this model.
The Hubble parameter scales as $H^2 \sim a^{2-3/n}$. It is then easy
to write it with its derivatives in terms of conformal
time,
 \be
   H^2 = \frac{4n^2}{(3-2n)^2}\frac{1}{\eta^2},
   \quad
   \dot{H} = \frac{1}{2}(2-\frac{3}{n})H^2, \quad
   \ddot{H} = \frac{1}{2}(2-\frac{3}{n})^2H^3.
 \ee
This solution applies also in the metric 
formulation\cite{Abdalla:2004sw}. In addition, 
Palatini variation of Ricci squared gravity, 
$f \sim (R^{\mu\nu}R_{\mu\nu})^{n/2}$, gives the same
expansion rate\cite{Allemandi:2004wn}.  

Here the scalar curvature is $R = 3(3-n)H^2/(2na^2)$. 
Since Eq.(\ref{dddelta}) involves only $\dot{F}/F$ and $\ddot{F}/F$, we
need only to know that
 \be
   \dot{F} = \frac{3(1-n)}{n}HF, \quad \ddot{F} =
   \frac{3(1-n)}{2n^2}(3-4n)H^2F.
 \ee
Substituting these in the general Eq.(\ref{dddelta}) yields
 \be
   (n-\frac{3}{2})\ddot{\delta}_v =  \frac{n}{\eta}\dot{\delta}_v +
   \left[-\frac{3}{\eta^2} + \frac{(n-1)(n-\frac{3}{2})}{3-n}k^2\right]\delta_v.
 \ee
Inserting a power law ansatz $\delta_v \sim a^m$ one finds that in
the large-scale limit $k=0$ the solutions are $m = -1/(6n)$ and $m =
3/n-2$. When $n$ is positive, the former is the decaying solution,
and the latter gives the growing mode, which reduces to the standard
matter-dominated solution $\delta \sim a$ when $n=1$. The growth rate
reduces when $n$ increases, reflecting the fact that the background
expansion is sped up. In the limit $n=3/2$, when $\ddot{a}=0$, then
also $\dot{\delta}_v=0$. For larger values of $n$ the
large-scale inhomogeneities in fact begin to smooth out, instead of
just their growth slowing down.

In the small-scale limit $k^2\eta^2 \gg 1$, and the gradient term
drives the perturbations to exponential growth
 \be
   \delta_v \sim \exp{\left(\pm  \sqrt{\frac{n-1}{3-n}}k\eta\right)},
 \ee
when $1>n>3$, or to oscillate,
 \be
   \delta_v \sim \exp{\left(\pm i\sqrt{\frac{n-1}{n-3}}k\eta\right)},
 \ee
when $n>3$. Since such features are not observed in the matter power
spectrum, these models do not appear to be viable alternatives to
dark energy. A quantitative study using existing observational data is 
under progress\cite{Koivisto:2006ie}.

\section{Discussion}

In the metric formulation nonlinear gravity theories result in rather
untractable fourth-order differential equations. In the Palatini
formulation the structure of the theory becomes interestingly
different. Since there is an algebraic relation between the curvature
scalar and the trace of the matter energy-momentum tensor, the
Palatini variation yields second-order field equations. In
Sec.~I we discussed observational and theoretical motivations for
studying these theories. In this paper we have demonstrated that even
the theory of cosmological perturbations is solvable for the Palatini
formulation of modified gravities.

In Sec.~II we presented  the equations for linear cosmological
perturbations in a general $f(R,\phi)$-gravity. We arranged them in a
gauge-ready form, from which one can easily adopt the most suitable
gauge for the particular problem at hand. We found that evolution of
vector and tensor perturbations are modestly modified from general
relativity, featuring some $f$-dependent prefactors. Field equations
for scalar modes of perturbations are modified in a more intricate
manner. In particular, nonlinear Palatini gravity features
nonstandard matter couplings, originating from perturbations in the
trace of the fluid energy momentum tensor and its derivatives. Only
in a specific gauge, $\delta T = 0$, these couplings ostensibly
disappear. These equations can be almost trivially generalized to
$f(R,\phi,X)$ gravity, where $X \equiv (\partial \phi)^2$. For
simplicity such a case was not included here, but curvature couplings
of scalar field derivatives have been introduced as an approach to
the cosmological constant problem\cite{Mukohyama:2003nw,
Dolgov:2003fw,Nojiri:2004bi, Nojiri:2004fw, Allemandi:2005qs}, and it
might be interesting to investigate implications of also these models
to cosmological perturbations.

In Sec.~III we derived also equations governing the growth of
inhomogeneities in the late universe undergoing an acceleration of
expansion driven by nonlinear $f(R)$-gravity. We found that an
effective pressure gradient term appears in the evolution equations
for cold dark matter inhomogeneities. This seems to be problematic
for structure formation in such models. Analytical solutions were
given for a simple example, $f(R) \sim R^n$, but the problem seems
generic to these nonlinear gravities. In the $f(R)=R-\alpha_0/(3R)$
case, Eq.(\ref{gradient2}) implies that the gradient term will drive
the perturbations at sufficiently small scales. It might
be difficult to tune the form of $f(R)$ (unless it is trivially close
to $f(R)=R-2\Lambda$) in such a way that the gradient term Eq.(\ref{gradient})
would be small enough inside the horizon not to affect the linear
matter power spectrum. However, an extensive study of the constraints
on the form and parameters of the function $f$ in the action
(\ref{action_p}) is left for forthcoming studies.

\appendix

\section{The equivalent metric formulation of a scalar-tensor theory}

In $n$ dimensions Eq.(\ref{correspondence}) generalizes to
 \be \label{correspondence2}
    \omega_{metric}(\phi) = \omega(\phi) -
    \frac{(n-1)}{(n-2)}\frac{F'^2(\phi)}{F(\phi)}.
 \ee
Let us therefore consider the action
 \be \label{action_mst}
 S = \int d^nx \sqrt{-g} \left\{\frac{1}{2}F(\phi) R(g) -
 \frac{1}{2}\left[\omega(\phi)-\frac{(n-1)}{(n-2)}\frac{F'(\phi)^2}{F(\phi)}\right]
     (\partial\phi)^2 - V(\phi) +
     \mathcal{L}_m(g_{\mu\nu},\Phi,...)\right\}.
 \ee
Variations of the metric lead to the field equations
 \bea
 F R^\mu_\nu(g) - \frac{1}{2}F R(g) & = & T^\mu_\nu
 +  (\nabla^\mu\nabla_\nu -\delta^\mu_\nu\Box )F
 \nonumber \\
 & + &
 \left[\omega-\frac{(n-1)}{(n-2)}\frac{F'^2}{F}\right](\nabla^\mu\phi)
 (\nabla_\nu\phi) -
 \delta^\mu_\nu\left[\left(\omega-\frac{(n-1)}{(n-2)}\frac{F'^2}{F}\right)
 (\partial\phi)^2
 + V\right],
 \eea
which can be rewritten as
 \bea \label{mstfield}
 G^\mu_\nu(g) & = & \frac{1}{F}\Bigg\{ T^\mu_\nu +
 T^{\mu (\phi)}_\nu
 \nonumber \\
 & + & \left(F''-\frac{(n-1)}{(n-2)}\frac{F'^2}{F}\right)
 (\nabla^\mu\phi)(\nabla_\nu\phi) +
 F'(\nabla^\mu\nabla_\nu\phi) +
 \delta^\mu_\nu\left[\left(-F''+\frac{(n-1)}{2(n-2)}\frac{F'^2}{F}\right)
 (\partial\phi)^2-
 F'\Box \phi\right]\Bigg\},
 \eea
where $T^{\mu (\phi)}_\nu$ is given by Eq.(\ref{scalaremt}).

Now let us vary the action of Eq.(\ref{action_p}) when $f = F(\phi)
R$. We then obtain the Ricci tensor from Eq.(\ref{riccit})
 \be
 R^\mu_\nu = R(g)^\mu_\nu + \frac{1}{F}\left\{
 \left[\frac{(n-1)}{(n-2)}\frac{F'^2}{F}-F''\right]
 (\nabla^\mu\phi)(\nabla_\nu\phi) +
 F'(\nabla^\mu\nabla_\nu\phi) -
 \delta^\mu_\nu\frac{1}{n-2}\left[F''(\partial
 \phi)^2 + F'\Box\phi\right]\right\},
 \ee
and the Ricci scalar from Eq.(\ref{riccis})
 \be
   R = R(g) +
   \frac{(n-1)}{(n-2)}\frac{1}{F}
   \left[\left(\frac{F'^2}{F}-2 F''\right)(\partial\phi)^2 - 2 F'\Box\phi\right].
 \ee
It is then straightforward to plug these into the field equation
(\ref{fields2}). The result is exactly Eq.(\ref{mstfield}), which
verifies that the scalar-tensor theory $F(\phi)R$ with the Palatini
variation is equivalent to the same theory with the metric variation
and the kinetic term rescaled according to
Eq.(\ref{correspondence2}). The equations of motion for the scalar
field are built into the field equations\cite{Koivisto:2005yk}, they 
also coincide in these equivalent theories.

\section{Scalar perturbations of the metric}

In this Appendix we present some formulae which have been used in the
derivation of the field equations. Here all covariant derivatives and
curvature variables correspond to the Levi-Civita connection of the
metric $g$ of Eq.(\ref{metric}). These results do not depend on the
variational principle which one uses. For a more extensive collection
of useful formulae for metric perturbations, see appendices in
Ref.\cite{Giovannini:2004rj}.

The Levi-Civita connection in the perturbed FRW spacetime,
Eq.(\ref{metric}), is
 \bea \label{toffels}
 \Gamma^0_{0 0} & = & H + \dot{\alpha}\,, \quad
 \Gamma^0_{0 i}   =   (\alpha-H\beta)_{,i}\,, \quad
 \Gamma^0_{ij}    =    g^{(3)}_{ij}\left[H(1-2\alpha+2\varphi)
                     +\dot{\varphi}\right] +
                 (2H\gamma+\dot{\gamma}+\beta)_{|ij}\,, \nonumber \\
 \Gamma^i_{0 0} & = & (\alpha-H\beta-\dot{\beta})^{,i}\,, \quad
 \Gamma^i_{0 j}   =    (H+\dot{\varphi})\delta^i_j +
  {{\dot{\gamma}}^{|i}}_{\phantom{|i}j} \,,
 \nonumber \\
 \Gamma^i_{jk}  & = & \Gamma^{i(3)}_{jk} + g^{(3)}_{jk}(H\beta - 
 \varphi)^{,i}
                  +\delta^i_j\varphi_{,k}+\delta^i_k\varphi_{,j} +
                {{\gamma_{|j}}^i}_k + {{\gamma_{|k}}^i}_j - 
 {{\gamma_{|j}}_k}^i \,.
  \eea
 Useful contractions of these are
  \be
  \Gamma^\mu_{\mu 0} = 4H + \dot{\alpha} + 3\dot{\varphi} +
            {{\dot{\gamma}}^{|k}}_{\phantom{|k} k} \,, \quad
  \Gamma^\mu_{\mu i} =  \Gamma^{k(3)}_{k i}  + \left[\alpha+3\varphi +
                   {\gamma^{|k}}_{k}\right]_{|i} \,.
  \ee
Covariant derivatives of a field $\xi (\bar{x},t)=\bar{\xi}(t)+\delta
 \xi(\bar{x},t)$ are then
  \bea
  a^2\nabla^0\nabla_0 \xi & = & -\ddot{\bar{\xi}} + H \dot{\bar{\xi}} - 
 \ddot{\delta \xi}
             + H\dot{\delta \xi} + \dot{\bar{\xi}}\dot{\alpha} +
               2\alpha(\ddot{\bar{\xi}} - H\dot{\bar{\xi}}) \,,
  \nonumber \\
  a^2\nabla^0\nabla_i \xi & = & \left[-\dot{\delta \xi}+ H\delta \xi +
                             \dot{\bar{\xi}}\alpha\right]_{,i} \,,
  \nonumber \\
  a^2\nabla^i\nabla_0 \xi & = & \left[\dot{\delta \xi} - H\delta \xi -
                             \ddot{\bar{\xi}}\beta +
                           \dot{\bar{\xi}}(\alpha+2H\beta)\right]^{,i} \,,
  \nonumber \\
  a^2\nabla^i\nabla_j \xi & = & -\delta^i_jH\dot{\bar{\xi}}
               +\left[\delta^i_j(2H\alpha-\dot{\varphi}) -
           \frac{1}{a}\chi^{|i}_{\phantom{|i}j}\right]\dot{\bar{\xi}}
               +\delta \xi^{|i}_{\phantom{|i}j} -  \delta^i_jH\dot{\delta 
 \xi} \,,
  \nonumber \\
  a^2\Box \xi & = & -\ddot{\bar{\xi}} - 2H\dot{\bar{\xi}} - \ddot{\delta 
 \xi}
             - 2H\dot{\delta \xi} +\delta \xi ^{|k}{\phantom{|k}k}
             + (2\ddot{\bar{\xi}} + H\dot{\bar{\xi}})\alpha
             + \dot{\bar{\xi}}(\dot{\alpha} + a\kappa) \,.
  \eea
 The Ricci tensor can be calculated using the ``hatless'' version of
 Eq.(\ref{ricci}),
  \be \label{riccihatless}
  R(g)_{\mu\nu} \equiv {\Gamma}^\alpha_{\mu\nu , \alpha}
        - {\Gamma}^\alpha_{\mu\alpha , \nu}
        + {\Gamma}^\alpha_{\alpha\lambda}{\Gamma}^\lambda_{\mu\nu}
         - {\Gamma}^\alpha_{\mu\lambda}{\Gamma}^\lambda_{\alpha\nu} \,,
  \ee
 with Eqs. (\ref{toffels}),
 \bea
  a^2R(g)^0_0 & = & 3\dot{H} - a\dot{\kappa}-2Ha\kappa + 
 3(H^2-\dot{H})\alpha
              -\triangle\alpha \,,
  \nonumber \\
  a^2R(g)^i_0 & = & 2\left[H\alpha
  -\dot{\varphi}+(\dot{H}-H^2)\beta\right]^{,i} \,,
  \nonumber \\
  a^2R(g)^0_i & = & 2\left[-H\alpha+\dot{\alpha}\right]_{,i}
              -   R(g)^{(3)}_{ik}\beta^{,i}, \,,
  \nonumber \\
  a^2R(g)^i_j & = & \delta^i_j(\dot{H}+2H^2)+R(g)^{i(3)}_j +
          \delta^i_j\left[\ddot{\varphi} -
          H(\dot{\alpha}-5\dot{\varphi}) -
          2(\dot{H}+2H^2)\alpha -2R^{i(3)}_j\varphi
          +  \Delta(-\varphi+\frac{H}{a}\chi)\right]
  \nonumber \\
          & + &
          \left[-\alpha-\varphi +          
 \frac{1}{a}\dot{\chi}+\frac{H}{a}\chi\right]^{|i}_{\phantom{|i}j}
          \,.
  \eea
 The curvature scalar is
  \be
  a^2R(g) =  6(\dot{H}+H^2) + R(g)^{(3)} - 2\left[ a\dot{\kappa} + 
 4Ha\kappa
         + 3(\dot{H}-H^2)\alpha + (R(g)^{(3)} + 2\triangle)\varphi +
         \triangle\alpha\right] \,.
  \ee

\acknowledgments{We thank Nikolay Koshelev, Tuomas Multam\"aki and Shinji Tsujikawa for useful comments.
TK is supported by the Magnus Ehrnrooth Foundation.}

\bibliography{refs2}

\begin{thebibliography}{75}
\expandafter\ifx\csname natexlab\endcsname\relax\def\natexlab#1{#1}\fi
\expandafter\ifx\csname bibnamefont\endcsname\relax
  \def\bibnamefont#1{#1}\fi
\expandafter\ifx\csname bibfnamefont\endcsname\relax
  \def\bibfnamefont#1{#1}\fi
\expandafter\ifx\csname citenamefont\endcsname\relax
  \def\citenamefont#1{#1}\fi
\expandafter\ifx\csname url\endcsname\relax
  \def\url#1{\texttt{#1}}\fi
\expandafter\ifx\csname urlprefix\endcsname\relax\def\urlprefix{URL }\fi
\providecommand{\bibinfo}[2]{#2}
\providecommand{\eprint}[2][]{\url{#2}}

\bibitem[{\citenamefont{Knop et~al.}(2003)}]{Knop:2003iy}
\bibinfo{author}{\bibfnamefont{R.~A.} \bibnamefont{Knop}} \bibnamefont{et~al.}
  (\bibinfo{collaboration}{The Supernova Cosmology Project}),
  \bibinfo{journal}{Astrophys. J.} \textbf{\bibinfo{volume}{598}},
  \bibinfo{pages}{102} (\bibinfo{year}{2003}), \eprint{astro-ph/0309368}.

\bibitem[{\citenamefont{Riess et~al.}(2004)}]{Riess:2004nr}
\bibinfo{author}{\bibfnamefont{A.~G.} \bibnamefont{Riess}} \bibnamefont{et~al.}
  (\bibinfo{collaboration}{Supernova Search Team}),
  \bibinfo{journal}{Astrophys. J.} \textbf{\bibinfo{volume}{607}},
  \bibinfo{pages}{665} (\bibinfo{year}{2004}), \eprint{astro-ph/0402512}.

\bibitem[{\citenamefont{Nesseris and Perivolaropoulos}(2004)}]{Nesseris:2004wj}
\bibinfo{author}{\bibfnamefont{S.}~\bibnamefont{Nesseris}} \bibnamefont{and}
  \bibinfo{author}{\bibfnamefont{L.}~\bibnamefont{Perivolaropoulos}},
  \bibinfo{journal}{Phys. Rev.} \textbf{\bibinfo{volume}{D70}},
  \bibinfo{pages}{043531} (\bibinfo{year}{2004}), \eprint{astro-ph/0401556}.

\bibitem[{\citenamefont{Spergel et~al.}(2003)}]{Spergel:2003cb}
\bibinfo{author}{\bibfnamefont{D.~N.} \bibnamefont{Spergel}}
  \bibnamefont{et~al.} (\bibinfo{collaboration}{WMAP}),
  \bibinfo{journal}{Astrophys. J. Suppl.} \textbf{\bibinfo{volume}{148}},
  \bibinfo{pages}{175} (\bibinfo{year}{2003}), \eprint{astro-ph/0302209}.

\bibitem[{\citenamefont{Carroll}(2001)}]{Carroll:2000fy}
\bibinfo{author}{\bibfnamefont{S.~M.} \bibnamefont{Carroll}},
  \bibinfo{journal}{Living Rev. Rel.} \textbf{\bibinfo{volume}{4}},
  \bibinfo{pages}{1} (\bibinfo{year}{2001}), \eprint{astro-ph/0004075}.

\bibitem[{\citenamefont{Peebles and Ratra}(2003)}]{Peebles:2002gy}
\bibinfo{author}{\bibfnamefont{P.~J.~E.} \bibnamefont{Peebles}}
  \bibnamefont{and} \bibinfo{author}{\bibfnamefont{B.}~\bibnamefont{Ratra}},
  \bibinfo{journal}{Rev. Mod. Phys.} \textbf{\bibinfo{volume}{75}},
  \bibinfo{pages}{559} (\bibinfo{year}{2003}), \eprint{astro-ph/0207347}.

\bibitem[{\citenamefont{Dvali et~al.}(2000)\citenamefont{Dvali, Gabadadze, and
  Porrati}}]{Dvali:2000hr}
\bibinfo{author}{\bibfnamefont{G.~R.} \bibnamefont{Dvali}},
  \bibinfo{author}{\bibfnamefont{G.}~\bibnamefont{Gabadadze}},
  \bibnamefont{and} \bibinfo{author}{\bibfnamefont{M.}~\bibnamefont{Porrati}},
  \bibinfo{journal}{Phys. Lett.} \textbf{\bibinfo{volume}{B485}},
  \bibinfo{pages}{208} (\bibinfo{year}{2000}), \eprint{hep-th/0005016}.

\bibitem[{\citenamefont{Arkani-Hamed et~al.}(2002)\citenamefont{Arkani-Hamed,
  Dimopoulos, Dvali, and Gabadadze}}]{Arkani-Hamed:2002fu}
\bibinfo{author}{\bibfnamefont{N.}~\bibnamefont{Arkani-Hamed}},
  \bibinfo{author}{\bibfnamefont{S.}~\bibnamefont{Dimopoulos}},
  \bibinfo{author}{\bibfnamefont{G.}~\bibnamefont{Dvali}}, \bibnamefont{and}
  \bibinfo{author}{\bibfnamefont{G.}~\bibnamefont{Gabadadze}}
  (\bibinfo{year}{2002}), \eprint{hep-th/0209227}.

\bibitem[{\citenamefont{Sahni and Shtanov}(2003)}]{Sahni:2002dx}
\bibinfo{author}{\bibfnamefont{V.}~\bibnamefont{Sahni}} \bibnamefont{and}
  \bibinfo{author}{\bibfnamefont{Y.}~\bibnamefont{Shtanov}},
  \bibinfo{journal}{JCAP} \textbf{\bibinfo{volume}{0311}}, \bibinfo{pages}{014}
  (\bibinfo{year}{2003}), \eprint{astro-ph/0202346}.

\bibitem[{\citenamefont{Dvali and Turner}(2003)}]{Dvali:2003rk}
\bibinfo{author}{\bibfnamefont{G.}~\bibnamefont{Dvali}} \bibnamefont{and}
  \bibinfo{author}{\bibfnamefont{M.~S.} \bibnamefont{Turner}}
  (\bibinfo{year}{2003}), \eprint{astro-ph/0301510}.

\bibitem[{\citenamefont{Cline and Vinet}(2003)}]{Cline:2002mw}
\bibinfo{author}{\bibfnamefont{J.~M.} \bibnamefont{Cline}} \bibnamefont{and}
  \bibinfo{author}{\bibfnamefont{J.}~\bibnamefont{Vinet}},
  \bibinfo{journal}{Phys. Rev.} \textbf{\bibinfo{volume}{D68}},
  \bibinfo{pages}{025015} (\bibinfo{year}{2003}), \eprint{hep-ph/0211284}.

\bibitem[{\citenamefont{Birrell and Davies}(1982)}]{Birrell:1982ix}
\bibinfo{author}{\bibfnamefont{N.~D.} \bibnamefont{Birrell}} \bibnamefont{and}
  \bibinfo{author}{\bibfnamefont{P.~C.~W.} \bibnamefont{Davies}},
  \emph{\bibinfo{title}{Quantum fields in curved space}}
  (\bibinfo{publisher}{Cambridge, UK: Univ. Pr.}, \bibinfo{year}{1982}),
  \bibinfo{note}{340p}.

\bibitem[{\citenamefont{Starobinsky}(1980)}]{Starobinsky:1980te}
\bibinfo{author}{\bibfnamefont{A.~A.} \bibnamefont{Starobinsky}},
  \bibinfo{journal}{Phys. Lett.} \textbf{\bibinfo{volume}{B91}},
  \bibinfo{pages}{99} (\bibinfo{year}{1980}).

\bibitem[{\citenamefont{Mukhanov et~al.}(1992)\citenamefont{Mukhanov, Feldman,
  and Brandenberger}}]{Mukhanov:1990me}
\bibinfo{author}{\bibfnamefont{V.~F.} \bibnamefont{Mukhanov}},
  \bibinfo{author}{\bibfnamefont{H.~A.} \bibnamefont{Feldman}},
  \bibnamefont{and} \bibinfo{author}{\bibfnamefont{R.~H.}
  \bibnamefont{Brandenberger}}, \bibinfo{journal}{Phys. Rept.}
  \textbf{\bibinfo{volume}{215}}, \bibinfo{pages}{203} (\bibinfo{year}{1992}).

\bibitem[{\citenamefont{Nojiri and
  Odintsov}(2003{\natexlab{a}})}]{Nojiri:2003ft}
\bibinfo{author}{\bibfnamefont{S.}~\bibnamefont{Nojiri}} \bibnamefont{and}
  \bibinfo{author}{\bibfnamefont{S.~D.} \bibnamefont{Odintsov}},
  \bibinfo{journal}{Phys. Rev.} \textbf{\bibinfo{volume}{D68}},
  \bibinfo{pages}{123512} (\bibinfo{year}{2003}{\natexlab{a}}),
  \eprint{hep-th/0307288}.

\bibitem[{\citenamefont{Capozziello et~al.}(2003)\citenamefont{Capozziello,
  Carloni, and Troisi}}]{Capozziello:2003tk}
\bibinfo{author}{\bibfnamefont{S.}~\bibnamefont{Capozziello}},
  \bibinfo{author}{\bibfnamefont{S.}~\bibnamefont{Carloni}}, \bibnamefont{and}
  \bibinfo{author}{\bibfnamefont{A.}~\bibnamefont{Troisi}}
  (\bibinfo{year}{2003}), \eprint{astro-ph/0303041}.

\bibitem[{\citenamefont{Carroll et~al.}(2004)\citenamefont{Carroll, Duvvuri,
  Trodden, and Turner}}]{Carroll:2003wy}
\bibinfo{author}{\bibfnamefont{S.~M.} \bibnamefont{Carroll}},
  \bibinfo{author}{\bibfnamefont{V.}~\bibnamefont{Duvvuri}},
  \bibinfo{author}{\bibfnamefont{M.}~\bibnamefont{Trodden}}, \bibnamefont{and}
  \bibinfo{author}{\bibfnamefont{M.~S.} \bibnamefont{Turner}},
  \bibinfo{journal}{Phys. Rev.} \textbf{\bibinfo{volume}{D70}},
  \bibinfo{pages}{043528} (\bibinfo{year}{2004}), \eprint{astro-ph/0306438}.

\bibitem[{\citenamefont{Nojiri and
  Odintsov}(2004{\natexlab{a}})}]{Nojiri:2003ni}
\bibinfo{author}{\bibfnamefont{S.}~\bibnamefont{Nojiri}} \bibnamefont{and}
  \bibinfo{author}{\bibfnamefont{S.~D.} \bibnamefont{Odintsov}},
  \bibinfo{journal}{Gen. Rel. Grav.} \textbf{\bibinfo{volume}{36}},
  \bibinfo{pages}{1765} (\bibinfo{year}{2004}{\natexlab{a}}),
  \eprint{hep-th/0308176}.

\bibitem[{\citenamefont{Nojiri and
  Odintsov}(2004{\natexlab{b}})}]{Nojiri:2004fw}
\bibinfo{author}{\bibfnamefont{S.}~\bibnamefont{Nojiri}} \bibnamefont{and}
  \bibinfo{author}{\bibfnamefont{S.~D.} \bibnamefont{Odintsov}},
  \bibinfo{journal}{Proc. Sci.} \textbf{\bibinfo{volume}{WC2004}},
  \bibinfo{pages}{024} (\bibinfo{year}{2004}{\natexlab{b}}),
  \eprint{hep-th/0412030}.

\bibitem[{\citenamefont{Nojiri and Odintsov}(2005)}]{Nojiri:2005jg}
\bibinfo{author}{\bibfnamefont{S.}~\bibnamefont{Nojiri}} \bibnamefont{and}
  \bibinfo{author}{\bibfnamefont{S.~D.} \bibnamefont{Odintsov}},
  \bibinfo{journal}{Phys. Lett.} \textbf{\bibinfo{volume}{B631}},
  \bibinfo{pages}{1} (\bibinfo{year}{2005}), \eprint{hep-th/0508049}.

\bibitem[{\citenamefont{Brevik et~al.}(2005)\citenamefont{Brevik, Gorbunova,
  and Shaido}}]{Brevik:2005ue}
\bibinfo{author}{\bibfnamefont{I.}~\bibnamefont{Brevik}},
  \bibinfo{author}{\bibfnamefont{O.}~\bibnamefont{Gorbunova}},
  \bibnamefont{and} \bibinfo{author}{\bibfnamefont{Y.~A.}
  \bibnamefont{Shaido}}, \bibinfo{journal}{Int. J. Mod. Phys.}
  \textbf{\bibinfo{volume}{D14}}, \bibinfo{pages}{1899} (\bibinfo{year}{2005}),
  \eprint{gr-qc/0508038}.

\bibitem[{\citenamefont{Multamaki and Vilja}(2006)}]{Multamaki:2005zs}
\bibinfo{author}{\bibfnamefont{T.}~\bibnamefont{Multamaki}} \bibnamefont{and}
  \bibinfo{author}{\bibfnamefont{I.}~\bibnamefont{Vilja}},
  \bibinfo{journal}{Phys. Rev.} \textbf{\bibinfo{volume}{D73}},
  \bibinfo{pages}{024018} (\bibinfo{year}{2006}), \eprint{astro-ph/0506692}.

\bibitem[{\citenamefont{Nojiri and
  Odintsov}(2003{\natexlab{b}})}]{Nojiri:2003rz}
\bibinfo{author}{\bibfnamefont{S.}~\bibnamefont{Nojiri}} \bibnamefont{and}
  \bibinfo{author}{\bibfnamefont{S.~D.} \bibnamefont{Odintsov}},
  \bibinfo{journal}{Phys. Lett.} \textbf{\bibinfo{volume}{B576}},
  \bibinfo{pages}{5} (\bibinfo{year}{2003}{\natexlab{b}}),
  \eprint{hep-th/0307071}.

\bibitem[{\citenamefont{Vollick}(2003)}]{Vollick:2003aw}
\bibinfo{author}{\bibfnamefont{D.~N.} \bibnamefont{Vollick}},
  \bibinfo{journal}{Phys. Rev.} \textbf{\bibinfo{volume}{D68}},
  \bibinfo{pages}{063510} (\bibinfo{year}{2003}), \eprint{astro-ph/0306630}.

\bibitem[{\citenamefont{Meng and Wang}(2003{\natexlab{a}})}]{Meng:2003ry}
\bibinfo{author}{\bibfnamefont{X.}~\bibnamefont{Meng}} \bibnamefont{and}
  \bibinfo{author}{\bibfnamefont{P.}~\bibnamefont{Wang}},
  \bibinfo{journal}{Class. Quant. Grav.} \textbf{\bibinfo{volume}{20}},
  \bibinfo{pages}{4949} (\bibinfo{year}{2003}{\natexlab{a}}),
  \eprint{astro-ph/0307354}.

\bibitem[{\citenamefont{Meng and Wang}(2004{\natexlab{a}})}]{Meng:2003uv}
\bibinfo{author}{\bibfnamefont{X.}~\bibnamefont{Meng}} \bibnamefont{and}
  \bibinfo{author}{\bibfnamefont{P.}~\bibnamefont{Wang}},
  \bibinfo{journal}{Class. Quant. Grav.} \textbf{\bibinfo{volume}{21}},
  \bibinfo{pages}{951} (\bibinfo{year}{2004}{\natexlab{a}}),
  \eprint{astro-ph/0308031}.

\bibitem[{\citenamefont{Misner et~al.}(1970)\citenamefont{Misner, Thorne, S.,
  and Wheeler}}]{Misner}
\bibinfo{author}{\bibfnamefont{C.}~\bibnamefont{Misner}},
  \bibinfo{author}{\bibnamefont{Thorne}},
  \bibinfo{author}{\bibfnamefont{K.}~\bibnamefont{S.}}, \bibnamefont{and}
  \bibinfo{author}{\bibfnamefont{J.}~\bibnamefont{Wheeler}},
  \emph{\bibinfo{title}{Gravitation}} (\bibinfo{publisher}{W.H. Freeman and
  Company}, \bibinfo{year}{1970}).

\bibitem[{\citenamefont{Smolin and Starodubtsev}(2003)}]{Smolin:2003qu}
\bibinfo{author}{\bibfnamefont{L.}~\bibnamefont{Smolin}} \bibnamefont{and}
  \bibinfo{author}{\bibfnamefont{A.}~\bibnamefont{Starodubtsev}}
  (\bibinfo{year}{2003}), \eprint{hep-th/0311163}.

\bibitem[{\citenamefont{Meng and Wang}(2004{\natexlab{b}})}]{Meng:2003sx}
\bibinfo{author}{\bibfnamefont{X.-H.} \bibnamefont{Meng}} \bibnamefont{and}
  \bibinfo{author}{\bibfnamefont{P.}~\bibnamefont{Wang}},
  \bibinfo{journal}{Gen. Rel. Grav.} \textbf{\bibinfo{volume}{36}},
  \bibinfo{pages}{1947} (\bibinfo{year}{2004}{\natexlab{b}}),
  \eprint{gr-qc/0311019}.

\bibitem[{\citenamefont{Dominguez and Barraco}(2004)}]{Dominguez:2004ds}
\bibinfo{author}{\bibfnamefont{A.~E.} \bibnamefont{Dominguez}}
  \bibnamefont{and} \bibinfo{author}{\bibfnamefont{D.~E.}
  \bibnamefont{Barraco}}, \bibinfo{journal}{Phys. Rev.}
  \textbf{\bibinfo{volume}{D70}}, \bibinfo{pages}{043505}
  (\bibinfo{year}{2004}), \eprint{gr-qc/0408069}.

\bibitem[{\citenamefont{Olmo}(2005)}]{Olmo:2005hd}
\bibinfo{author}{\bibfnamefont{G.~J.} \bibnamefont{Olmo}}
  (\bibinfo{year}{2005}), \eprint{gr-qc/0505136}.

\bibitem[{\citenamefont{Allemandi
  et~al.}(2005{\natexlab{a}})\citenamefont{Allemandi, Francaviglia, Ruggiero,
  and Tartaglia}}]{Allemandi:2005tg}
\bibinfo{author}{\bibfnamefont{G.}~\bibnamefont{Allemandi}},
  \bibinfo{author}{\bibfnamefont{M.}~\bibnamefont{Francaviglia}},
  \bibinfo{author}{\bibfnamefont{M.~L.} \bibnamefont{Ruggiero}},
  \bibnamefont{and}
  \bibinfo{author}{\bibfnamefont{A.}~\bibnamefont{Tartaglia}},
  \bibinfo{journal}{Gen. Rel. Grav.} \textbf{\bibinfo{volume}{37}},
  \bibinfo{pages}{1891} (\bibinfo{year}{2005}{\natexlab{a}}),
  \eprint{gr-qc/0506123}.

\bibitem[{\citenamefont{Sotiriou}(2005)}]{Sotiriou:2005xe}
\bibinfo{author}{\bibfnamefont{T.~P.} \bibnamefont{Sotiriou}}
  (\bibinfo{year}{2005}), \eprint{gr-qc/0507027}.

\bibitem[{\citenamefont{Sotiriou}(2006)}]{Sotiriou:2005hu}
\bibinfo{author}{\bibfnamefont{T.~P.} \bibnamefont{Sotiriou}},
  \bibinfo{journal}{Phys. Rev.} \textbf{\bibinfo{volume}{D73}},
  \bibinfo{pages}{063515} (\bibinfo{year}{2006}), \eprint{gr-qc/0509029}.

\bibitem[{\citenamefont{Allemandi
  et~al.}(2004{\natexlab{a}})\citenamefont{Allemandi, Borowiec, and
  Francaviglia}}]{Allemandi:2004ca}
\bibinfo{author}{\bibfnamefont{G.}~\bibnamefont{Allemandi}},
  \bibinfo{author}{\bibfnamefont{A.}~\bibnamefont{Borowiec}}, \bibnamefont{and}
  \bibinfo{author}{\bibfnamefont{M.}~\bibnamefont{Francaviglia}},
  \bibinfo{journal}{Phys. Rev.} \textbf{\bibinfo{volume}{D70}},
  \bibinfo{pages}{043524} (\bibinfo{year}{2004}{\natexlab{a}}),
  \eprint{hep-th/0403264}.

\bibitem[{\citenamefont{Meng and Wang}(2004{\natexlab{c}})}]{Meng:2003en}
\bibinfo{author}{\bibfnamefont{X.-H.} \bibnamefont{Meng}} \bibnamefont{and}
  \bibinfo{author}{\bibfnamefont{P.}~\bibnamefont{Wang}},
  \bibinfo{journal}{Phys. Lett.} \textbf{\bibinfo{volume}{B584}},
  \bibinfo{pages}{1} (\bibinfo{year}{2004}{\natexlab{c}}),
  \eprint{hep-th/0309062}.

\bibitem[{\citenamefont{Vollick}(2004)}]{Vollick:2003ic}
\bibinfo{author}{\bibfnamefont{D.~N.} \bibnamefont{Vollick}},
  \bibinfo{journal}{Class. Quant. Grav.} \textbf{\bibinfo{volume}{21}},
  \bibinfo{pages}{3813} (\bibinfo{year}{2004}), \eprint{gr-qc/0312041}.

\bibitem[{\citenamefont{Meng and Wang}(2003{\natexlab{b}})}]{Meng:2003bk}
\bibinfo{author}{\bibfnamefont{X.-H.} \bibnamefont{Meng}} \bibnamefont{and}
  \bibinfo{author}{\bibfnamefont{P.}~\bibnamefont{Wang}}
  (\bibinfo{year}{2003}{\natexlab{b}}), \eprint{astro-ph/0308284}.

\bibitem[{\citenamefont{Meng and Wang}(2005)}]{Meng:2004wg}
\bibinfo{author}{\bibfnamefont{X.~H.} \bibnamefont{Meng}} \bibnamefont{and}
  \bibinfo{author}{\bibfnamefont{P.}~\bibnamefont{Wang}},
  \bibinfo{journal}{Class. Quant. Grav.} \textbf{\bibinfo{volume}{22}},
  \bibinfo{pages}{23} (\bibinfo{year}{2005}), \eprint{gr-qc/0411007}.

\bibitem[{\citenamefont{Kremer and Alves}(2004)}]{Kremer:2004bf}
\bibinfo{author}{\bibfnamefont{G.~M.} \bibnamefont{Kremer}} \bibnamefont{and}
  \bibinfo{author}{\bibfnamefont{D.~S.~M.} \bibnamefont{Alves}},
  \bibinfo{journal}{Phys. Rev.} \textbf{\bibinfo{volume}{D70}},
  \bibinfo{pages}{023503} (\bibinfo{year}{2004}), \eprint{gr-qc/0404082}.

\bibitem[{\citenamefont{Capozziello et~al.}(2004)\citenamefont{Capozziello,
  Cardone, and Francaviglia}}]{Capozziello:2004vh}
\bibinfo{author}{\bibfnamefont{S.}~\bibnamefont{Capozziello}},
  \bibinfo{author}{\bibfnamefont{V.~F.} \bibnamefont{Cardone}},
  \bibnamefont{and}
  \bibinfo{author}{\bibfnamefont{M.}~\bibnamefont{Francaviglia}}
  (\bibinfo{year}{2004}), \eprint{astro-ph/0410135}.

\bibitem[{\citenamefont{Lue et~al.}(2004)\citenamefont{Lue, Scoccimarro, and
  Starkman}}]{Lue:2003ky}
\bibinfo{author}{\bibfnamefont{A.}~\bibnamefont{Lue}},
  \bibinfo{author}{\bibfnamefont{R.}~\bibnamefont{Scoccimarro}},
  \bibnamefont{and} \bibinfo{author}{\bibfnamefont{G.}~\bibnamefont{Starkman}},
  \bibinfo{journal}{Phys. Rev.} \textbf{\bibinfo{volume}{D69}},
  \bibinfo{pages}{044005} (\bibinfo{year}{2004}), \eprint{astro-ph/0307034}.

\bibitem[{\citenamefont{Koivisto et~al.}(2005)\citenamefont{Koivisto,
  Kurki-Suonio, and Ravndal}}]{Koivisto:2004ne}
\bibinfo{author}{\bibfnamefont{T.}~\bibnamefont{Koivisto}},
  \bibinfo{author}{\bibfnamefont{H.}~\bibnamefont{Kurki-Suonio}},
  \bibnamefont{and} \bibinfo{author}{\bibfnamefont{F.}~\bibnamefont{Ravndal}},
  \bibinfo{journal}{Phys. Rev.} \textbf{\bibinfo{volume}{D71}},
  \bibinfo{pages}{064027} (\bibinfo{year}{2005}), \eprint{astro-ph/0409163}.

\bibitem[{\citenamefont{Koivisto}(2005{\natexlab{a}})}]{Koivisto:2005nr}
\bibinfo{author}{\bibfnamefont{T.}~\bibnamefont{Koivisto}},
  \bibinfo{journal}{Phys. Rev.} \textbf{\bibinfo{volume}{D72}},
  \bibinfo{pages}{043516} (\bibinfo{year}{2005}{\natexlab{a}}),
  \eprint{astro-ph/0504571}.

\bibitem[{\citenamefont{Linder}(2005)}]{Linder:2005in}
\bibinfo{author}{\bibfnamefont{E.~V.} \bibnamefont{Linder}},
  \bibinfo{journal}{Phys. Rev.} \textbf{\bibinfo{volume}{D72}},
  \bibinfo{pages}{043529} (\bibinfo{year}{2005}), \eprint{astro-ph/0507263}.

\bibitem[{\citenamefont{Bardeen}(1980)}]{Bardeen:1980kt}
\bibinfo{author}{\bibfnamefont{J.~M.} \bibnamefont{Bardeen}},
  \bibinfo{journal}{Phys. Rev.} \textbf{\bibinfo{volume}{D22}},
  \bibinfo{pages}{1882} (\bibinfo{year}{1980}).

\bibitem[{\citenamefont{Ellis and Bruni}(1989)}]{Ellis:1989jt}
\bibinfo{author}{\bibfnamefont{G.~F.~R.} \bibnamefont{Ellis}} \bibnamefont{and}
  \bibinfo{author}{\bibfnamefont{M.}~\bibnamefont{Bruni}},
  \bibinfo{journal}{Phys. Rev.} \textbf{\bibinfo{volume}{D40}},
  \bibinfo{pages}{1804} (\bibinfo{year}{1989}).

\bibitem[{\citenamefont{Ellis et~al.}(1989)\citenamefont{Ellis, Hwang, and
  Bruni}}]{Ellis:1989ju}
\bibinfo{author}{\bibfnamefont{G.~F.~R.} \bibnamefont{Ellis}},
  \bibinfo{author}{\bibfnamefont{J.}~\bibnamefont{Hwang}}, \bibnamefont{and}
  \bibinfo{author}{\bibfnamefont{M.}~\bibnamefont{Bruni}},
  \bibinfo{journal}{Phys. Rev.} \textbf{\bibinfo{volume}{D40}},
  \bibinfo{pages}{1819} (\bibinfo{year}{1989}).

\bibitem[{\citenamefont{Hwang}(1990)}]{Hwang:1990re}
\bibinfo{author}{\bibfnamefont{J.~C.} \bibnamefont{Hwang}},
  \bibinfo{journal}{Class. Quant. Grav.} \textbf{\bibinfo{volume}{7}},
  \bibinfo{pages}{1613} (\bibinfo{year}{1990}).

\bibitem[{\citenamefont{Hwang}(1991)}]{Hwang:1991aj}
\bibinfo{author}{\bibfnamefont{J.-c.} \bibnamefont{Hwang}},
  \bibinfo{journal}{Astrophys. J.} \textbf{\bibinfo{volume}{375}},
  \bibinfo{pages}{443} (\bibinfo{year}{1991}).

\bibitem[{\citenamefont{Hwang and Noh}(1996)}]{Hwang:1996xh}
\bibinfo{author}{\bibfnamefont{J.-c.} \bibnamefont{Hwang}} \bibnamefont{and}
  \bibinfo{author}{\bibfnamefont{H.}~\bibnamefont{Noh}},
  \bibinfo{journal}{Phys. Rev.} \textbf{\bibinfo{volume}{D54}},
  \bibinfo{pages}{1460} (\bibinfo{year}{1996}).

\bibitem[{\citenamefont{Hwang and Noh}(2002{\natexlab{a}})}]{Hwang:2001qk}
\bibinfo{author}{\bibfnamefont{J.-c.} \bibnamefont{Hwang}} \bibnamefont{and}
  \bibinfo{author}{\bibfnamefont{H.-r.} \bibnamefont{Noh}},
  \bibinfo{journal}{Phys. Rev.} \textbf{\bibinfo{volume}{D65}},
  \bibinfo{pages}{023512} (\bibinfo{year}{2002}{\natexlab{a}}),
  \eprint{astro-ph/0102005}.

\bibitem[{\citenamefont{Hwang and Noh}(2002{\natexlab{b}})}]{Hwang:2002fp}
\bibinfo{author}{\bibfnamefont{J.-c.} \bibnamefont{Hwang}} \bibnamefont{and}
  \bibinfo{author}{\bibfnamefont{H.}~\bibnamefont{Noh}},
  \bibinfo{journal}{Phys. Rev.} \textbf{\bibinfo{volume}{D66}},
  \bibinfo{pages}{084009} (\bibinfo{year}{2002}{\natexlab{b}}),
  \eprint{hep-th/0206100}.

\bibitem[{\citenamefont{Hwang and Noh}(2005)}]{Hwang:2005hb}
\bibinfo{author}{\bibfnamefont{J.-c.} \bibnamefont{Hwang}} \bibnamefont{and}
  \bibinfo{author}{\bibfnamefont{H.}~\bibnamefont{Noh}},
  \bibinfo{journal}{Phys. Rev.} \textbf{\bibinfo{volume}{D71}},
  \bibinfo{pages}{063536} (\bibinfo{year}{2005}), \eprint{gr-qc/0412126}.

\bibitem[{\citenamefont{Koivisto}(2005{\natexlab{b}})}]{Koivisto:2005yk}
\bibinfo{author}{\bibfnamefont{T.}~\bibnamefont{Koivisto}}
  (\bibinfo{year}{2005}{\natexlab{b}}), \eprint{gr-qc/0505128}.

\bibitem[{\citenamefont{Magnano}(1995)}]{Magnano:1995pv}
\bibinfo{author}{\bibfnamefont{G.}~\bibnamefont{Magnano}}
  (\bibinfo{year}{1995}), \eprint{gr-qc/9511027}.

\bibitem[{\citenamefont{Ferraris et~al.}(1992)\citenamefont{Ferraris,
  Francaviglia, and Volovich}}]{Ferraris:1992dx}
\bibinfo{author}{\bibfnamefont{M.}~\bibnamefont{Ferraris}},
  \bibinfo{author}{\bibfnamefont{M.}~\bibnamefont{Francaviglia}},
  \bibnamefont{and} \bibinfo{author}{\bibfnamefont{I.}~\bibnamefont{Volovich}}
  (\bibinfo{year}{1992}), \eprint{gr-qc/9303007}.

\bibitem[{\citenamefont{Alnes et~al.}(2005)\citenamefont{Alnes, Ravndal, and
  Wehus}}]{Alnes:2005ed}
\bibinfo{author}{\bibfnamefont{H.}~\bibnamefont{Alnes}},
  \bibinfo{author}{\bibfnamefont{F.}~\bibnamefont{Ravndal}}, \bibnamefont{and}
  \bibinfo{author}{\bibfnamefont{I.~K.} \bibnamefont{Wehus}}
  (\bibinfo{year}{2005}), \eprint{quant-ph/0506131}.

\bibitem[{\citenamefont{Allemandi et~al.}(2006)\citenamefont{Allemandi, Capone,
  Capozziello, and Francaviglia}}]{Allemandi:2004yx}
\bibinfo{author}{\bibfnamefont{G.}~\bibnamefont{Allemandi}},
  \bibinfo{author}{\bibfnamefont{M.}~\bibnamefont{Capone}},
  \bibinfo{author}{\bibfnamefont{S.}~\bibnamefont{Capozziello}},
  \bibnamefont{and}
  \bibinfo{author}{\bibfnamefont{M.}~\bibnamefont{Francaviglia}},
  \bibinfo{journal}{Gen. Rel. Grav.} \textbf{\bibinfo{volume}{38}},
  \bibinfo{pages}{33} (\bibinfo{year}{2006}), \eprint{hep-th/0409198}.

\bibitem[{\citenamefont{Allemandi
  et~al.}(2005{\natexlab{b}})\citenamefont{Allemandi, Borowiec, Francaviglia,
  and Odintsov}}]{Allemandi:2005qs}
\bibinfo{author}{\bibfnamefont{G.}~\bibnamefont{Allemandi}},
  \bibinfo{author}{\bibfnamefont{A.}~\bibnamefont{Borowiec}},
  \bibinfo{author}{\bibfnamefont{M.}~\bibnamefont{Francaviglia}},
  \bibnamefont{and} \bibinfo{author}{\bibfnamefont{S.~D.}
  \bibnamefont{Odintsov}}, \bibinfo{journal}{Phys. Rev.}
  \textbf{\bibinfo{volume}{D72}}, \bibinfo{pages}{063505}
  (\bibinfo{year}{2005}{\natexlab{b}}), \eprint{gr-qc/0504057}.

\bibitem[{\citenamefont{Fujii and Maeda}(2003)}]{Fujii:2003pa}
\bibinfo{author}{\bibfnamefont{Y.}~\bibnamefont{Fujii}} \bibnamefont{and}
  \bibinfo{author}{\bibfnamefont{K.}~\bibnamefont{Maeda}},
  \emph{\bibinfo{title}{The scalar-tensor theory of gravitation}}
  (\bibinfo{publisher}{Cambridge, USA: Univ. Pr.}, \bibinfo{year}{2003}),
  \bibinfo{note}{240 p}.

\bibitem[{\citenamefont{Malik and Wands}(2005)}]{Malik:2004tf}
\bibinfo{author}{\bibfnamefont{K.~A.} \bibnamefont{Malik}} \bibnamefont{and}
  \bibinfo{author}{\bibfnamefont{D.}~\bibnamefont{Wands}},
  \bibinfo{journal}{JCAP} \textbf{\bibinfo{volume}{0502}}, \bibinfo{pages}{007}
  (\bibinfo{year}{2005}), \eprint{astro-ph/0411703}.

\bibitem[{\citenamefont{Wang and Meng}(2004)}]{Wang:2004vs}
\bibinfo{author}{\bibfnamefont{P.}~\bibnamefont{Wang}} \bibnamefont{and}
  \bibinfo{author}{\bibfnamefont{X.-H.} \bibnamefont{Meng}},
  \bibinfo{journal}{TSPU Vestnik} \textbf{\bibinfo{volume}{44N7}},
  \bibinfo{pages}{40} (\bibinfo{year}{2004}), \eprint{astro-ph/0406455}.

\bibitem[{\citenamefont{Amarzguioui et~al.}(2004)\citenamefont{Amarzguioui,
  Elgaroy, and Multamaki}}]{Amarzguioui:2004kc}
\bibinfo{author}{\bibfnamefont{M.}~\bibnamefont{Amarzguioui}},
  \bibinfo{author}{\bibfnamefont{O.}~\bibnamefont{Elgaroy}}, \bibnamefont{and}
  \bibinfo{author}{\bibfnamefont{T.}~\bibnamefont{Multamaki}}
  (\bibinfo{year}{2004}), \eprint{astro-ph/0410408}.

\bibitem[{\citenamefont{Dolgov and
  Kawasaki}(2003{\natexlab{a}})}]{Dolgov:2003px}
\bibinfo{author}{\bibfnamefont{A.~D.} \bibnamefont{Dolgov}} \bibnamefont{and}
  \bibinfo{author}{\bibfnamefont{M.}~\bibnamefont{Kawasaki}},
  \bibinfo{journal}{Phys. Lett.} \textbf{\bibinfo{volume}{B573}},
  \bibinfo{pages}{1} (\bibinfo{year}{2003}{\natexlab{a}}),
  \eprint{astro-ph/0307285}.

\bibitem[{\citenamefont{Nojiri and
  Odintsov}(2004{\natexlab{c}})}]{Nojiri:2003wx}
\bibinfo{author}{\bibfnamefont{S.}~\bibnamefont{Nojiri}} \bibnamefont{and}
  \bibinfo{author}{\bibfnamefont{S.~D.} \bibnamefont{Odintsov}},
  \bibinfo{journal}{Mod. Phys. Lett.} \textbf{\bibinfo{volume}{A19}},
  \bibinfo{pages}{627} (\bibinfo{year}{2004}{\natexlab{c}}),
  \eprint{hep-th/0310045}.

\bibitem[{\citenamefont{Faraoni}(2004)}]{Faraoni:2004dn}
\bibinfo{author}{\bibfnamefont{V.}~\bibnamefont{Faraoni}},
  \bibinfo{journal}{Phys. Rev.} \textbf{\bibinfo{volume}{D70}},
  \bibinfo{pages}{044037} (\bibinfo{year}{2004}), \eprint{gr-qc/0407021}.

\bibitem[{\citenamefont{Faraoni}(2005)}]{Faraoni:2005ie}
\bibinfo{author}{\bibfnamefont{V.}~\bibnamefont{Faraoni}}
  (\bibinfo{year}{2005}), \eprint{gr-qc/0509008}.

\bibitem[{\citenamefont{Allemandi
  et~al.}(2004{\natexlab{b}})\citenamefont{Allemandi, Borowiec, and
  Francaviglia}}]{Allemandi:2004wn}
\bibinfo{author}{\bibfnamefont{G.}~\bibnamefont{Allemandi}},
  \bibinfo{author}{\bibfnamefont{A.}~\bibnamefont{Borowiec}}, \bibnamefont{and}
  \bibinfo{author}{\bibfnamefont{M.}~\bibnamefont{Francaviglia}},
  \bibinfo{journal}{Phys. Rev.} \textbf{\bibinfo{volume}{D70}},
  \bibinfo{pages}{103503} (\bibinfo{year}{2004}{\natexlab{b}}),
  \eprint{hep-th/0407090}.

\bibitem[{\citenamefont{Abdalla et~al.}(2005)\citenamefont{Abdalla, Nojiri, and
  Odintsov}}]{Abdalla:2004sw}
\bibinfo{author}{\bibfnamefont{M.~C.~B.} \bibnamefont{Abdalla}},
  \bibinfo{author}{\bibfnamefont{S.}~\bibnamefont{Nojiri}}, \bibnamefont{and}
  \bibinfo{author}{\bibfnamefont{S.~D.} \bibnamefont{Odintsov}},
  \bibinfo{journal}{Class. Quant. Grav.} \textbf{\bibinfo{volume}{22}},
  \bibinfo{pages}{L35} (\bibinfo{year}{2005}), \eprint{hep-th/0409177}.

\bibitem[{\citenamefont{Koivisto}(2006)}]{Koivisto:2006ie}
\bibinfo{author}{\bibfnamefont{T.}~\bibnamefont{Koivisto}},
  \bibinfo{journal}{Phys. Rev.} \textbf{\bibinfo{volume}{D73}},
  \bibinfo{pages}{083517} (\bibinfo{year}{2006}), \eprint{astro-ph/0602031}.

\bibitem[{\citenamefont{Mukohyama and Randall}(2004)}]{Mukohyama:2003nw}
\bibinfo{author}{\bibfnamefont{S.}~\bibnamefont{Mukohyama}} \bibnamefont{and}
  \bibinfo{author}{\bibfnamefont{L.}~\bibnamefont{Randall}},
  \bibinfo{journal}{Phys. Rev. Lett.} \textbf{\bibinfo{volume}{92}},
  \bibinfo{pages}{211302} (\bibinfo{year}{2004}), \eprint{hep-th/0306108}.

\bibitem[{\citenamefont{Dolgov and
  Kawasaki}(2003{\natexlab{b}})}]{Dolgov:2003fw}
\bibinfo{author}{\bibfnamefont{A.~D.} \bibnamefont{Dolgov}} \bibnamefont{and}
  \bibinfo{author}{\bibfnamefont{M.}~\bibnamefont{Kawasaki}}
  (\bibinfo{year}{2003}{\natexlab{b}}), \eprint{astro-ph/0307442}.

\bibitem[{\citenamefont{Nojiri and
  Odintsov}(2004{\natexlab{d}})}]{Nojiri:2004bi}
\bibinfo{author}{\bibfnamefont{S.}~\bibnamefont{Nojiri}} \bibnamefont{and}
  \bibinfo{author}{\bibfnamefont{S.~D.} \bibnamefont{Odintsov}},
  \bibinfo{journal}{Phys. Lett.} \textbf{\bibinfo{volume}{B599}},
  \bibinfo{pages}{137} (\bibinfo{year}{2004}{\natexlab{d}}),
  \eprint{astro-ph/0403622}.

\bibitem[{\citenamefont{Giovannini}(2005)}]{Giovannini:2004rj}
\bibinfo{author}{\bibfnamefont{M.}~\bibnamefont{Giovannini}},
  \bibinfo{journal}{Int. J. Mod. Phys.} \textbf{\bibinfo{volume}{D14}},
  \bibinfo{pages}{363} (\bibinfo{year}{2005}), \eprint{astro-ph/0412601}.

\end{thebibliography}

\end{document}